\documentclass[12pt]{article} 
\begin{document} 
\title{\vskip -5cm {\small\begin{flushright} UMH-MG-98/03\\
ULB-TH-98/12\end{flushright}}
\vskip 2.5 cm 
{\bf Mimimal Length Uncertainty Principle and the Transplanckian
Problem of Black Hole Physics}}
\author{R. Brout \thanks{E-mail: rbrout@mach.ulb.ac.be}$\mbox{}^{\mbox{\footnotesize ,\ a}}$
\and 
Cl. Gabriel\thanks{E-mail gabriel@sun1.umh.ac.be}$\mbox{}^{\mbox{\footnotesize ,\ b}}$
\and 
M. Lubo\thanks{E-mail lubo@sun1.umh.ac.be}$\mbox{}^{\mbox{\footnotesize ,\ b}}$
\and 
Ph. Spindel\thanks{E-mail spindel@sun1.umh.ac.be}$\mbox{}^{\mbox{\footnotesize ,\ b}}$} 
\date{
 a) {\em
Service de Physique th\'eorique, Universit\'e Libre de Bruxelles}, \\ {\em  C. P. 225, bvd du
Triomphe, B-1050 Bruxelles, Belgium} \\
b) {\em M\'ecanique et Gravitation,
Universit\'e de Mons-Hainaut}, \\ {\em 6, avenue du Champ de Mars, B-7000 Mons, Belgium}}
\maketitle
\begin{abstract} The minimal length uncertainty principle of Kempf,
Mangano and Mann (KMM), as derived from a mutilated quantum commutator between coordinate and momentum, is
applied to describe the modes and wave packets of Hawking particles evaporated from a black hole. The
transplanckian problem is successfully confronted in that the Hawking particle no longer hugs the
horizon at arbitrarily close distances. Rather the mode of Schwarzschild frequency $\omega$ deviates
from the conventional trajectory when the coordinate $r$ is given by $\vert r - 2M\vert\simeq \beta_H
\omega / 2 \pi$ in units of the non local distance legislated into the uncertainty relation.Wave
packets straddle the horizon and spread out to fill the whole non local region. The charge carried by
the packet (in the sense of the amount of "stuff" carried by the Klein--Gordon field) is not
conserved in the non--local region and rapidly decreases to zero as time decreases. Read in the forward
temporal direction, the non--local region thus is the seat of production of the Hawking particle and
its partner. The KMM model was inspired by string theory for which the mutilated commutator has been
proposed to describe an effective theory of high momentum scattering of zero mass modes. It is here
interpreted in terms of dissipation which gives rise to the Hawking particle into a reservoir of
other modes (of as yet unknown origin). On this basis it is conjectured that the Bekenstein--Hawking
entropy finds its origin in the fluctuations of fields extending over the non local
region.\end{abstract}
\vfill
\newpage
\section{Introduction}
  More than two decades after its discovery,  Hawking's theory of 
black hole evaporation  \cite{Hawk} continues to be plagued by the "transplanckian"
problem. Presumably,  due to an insufficient treatment of the gravitational back
reaction,  the theory suffers from an overdose of localization. Outgoing photons
hug the horizon ($r= 2M$ in planckian units at a distance $O(M e^{-\alpha
M^2})$ with $\alpha= O(1)$). This results in proper energies near the horizon
$O(\omega e^{\alpha M^2})$ where $\omega$ is the observed energy at asymptotic
distances; typically $\omega = O(M^{-1} )$ and $M= O(10^{40})$ for a macroscopic
black hole. \\
   Considerable effort has gone into introducing the necessary ingredients of
non--locality to cure this disease. Relevant to the present paper is the
methodology \cite{BMPS} that has been brought to bear to  exploit Unruh's
 model of the dumb hole \cite{Unruh}. In particular,  we shall use
Eddington--Finkelstein coordinates (EF) as introduced by Damour--Ruffini (DR). 
This turns out to be an efficient tool. Though interesting in itself,  in that
the dumb hole analogy shows the robustness of Hawking's radiation in resisting
mutilation of the conventional theory,  it is nevertheless inadequate. This is
because in fluids,  there is a cut--off in momentum as well as 
in energy. In adopting this to the black hole problem, as in ref. \cite{BMPS}, one cuts off the energy but not the
momentum,  elsewise one would lose the Hawking effect. There has, as yet,  been
no justification offered for this procedure.\\
  In this paper,  using the DR technique,  we introduce an alternative
mechanism of non--locality which we believe has some chance of being  
founded in the correct physics of the situation. This is the non--local commutator,  the
object of study of Kempf,  Mangano and Mann (KMM) \cite{KMM}. 
 Such "mutilated" commutators have been proposed in the context of string  
theory (see refs \cite{KMM} and \cite{Kempf} for a bibliography),  but 
they may arise in a more general
context wherein the mode relevant to a particular
 problem (like the evaporating photon) interacts with "reservoir" modes in
general. In this paper we do not enter into this fundamental question  
aside from some (superficial) concluding remarks. Rather we take the pragmatic point
of view : assume KMM and see if it cures the transplanckian  
problem. And it does ! Furthermore,  in so doing the formalism suggests the origin of the
Bekenstein--Hawking black hole entropy \cite{Bek},  as we shall point out at the
end of the development. The main physical picture that emerges is that the
reservoir,  which is responsible for KMM non--locality,  boils off a Hawking pair
on either side of the horizon,  in a region whose extension is $(\beta_H
\omega)$ units of the non local length scale. 
 Here $\beta_H= 8 \pi M$ is the inverse Hawking temperature. 
\section{KMM theory}
  We begin with a brief summary of KMM. The point of departure is
 (for one degree of freedom)
\begin{equation}
[\hat P,\hat Q]= -i(1+\frac {\hat P^2}{\mu^2}) \label{1}
\end{equation}
In what follows we adopt for momentum and coordinate the non dimensional 
variables $\hat p= \hat P/ \mu$; $\hat q= \mu\; \hat Q$. The scale $\mu$ could be planckian or  
it
could involve some fractional power of $M$. All we require is $(M/m_{pl}^2)>> \mu
^{-1}$ in order to have a sufficiently large asymptotic region so that
there exists a region $\mu^{-1}<<(r-2M)<<2M$; from now on all lengths are in Planckian units. \\
  Equation (\ref{1}) implies a modified uncertainty relation
\begin{equation}
\Delta p \;\Delta q \geq \frac 12 [1+ \langle \hat p^2 \rangle]= \frac 12 [1+(\Delta
p)^2+\langle \hat p\rangle ^2] \label{2}
\end{equation}
where $\Delta p \equiv \langle (\Delta \hat p)^2 \rangle ^{1/2}, \Delta q 
\equiv \langle (\Delta \hat q)^2 \rangle ^{1/2}$. Thus $\Delta q$ has a
minimal value (= unity) at $\langle \hat p \rangle= 0$ and $\Delta p= 1$. 
We note in passing the frame dependence. This has not been studied
either by KMM or by us. Perhaps its elucidation will require a careful
study of the underlying fundamental mechanisms behind eq. (\ref{1}). In
keeping with this,  our initial pragmatic exploration,  we adopt the most
natural assumption. As
 in ref \cite{BMPS},  the frame is taken to be the rest frame of the black hole. \\
The main thrust of KMM is the search for a Hilbert space formalism
which is physically sensible. Thus one requires that all expectation values of
$\hat q$ be real. The matter is subtle in that there are no eigenstates of
$\hat q$ compatible with eq. (\ref{2}) whereas eigenstates of $\hat p$ do exist. 
So one works in $p$ representation and requires that $
\hat q$ be a symmetric operator. The scalar product of two state vectors, $\vert f
\rangle$ and $\vert g \rangle$, is thus conveniently represented by
\begin{equation}
\langle f, g\rangle= \int_\infty ^\infty  f^*(p) g(p)\, \frac {dp} {1+p^2} =
\int_{\pi /2}^{\pi /2} f^*(\tan \theta) g(\tan \theta)\,  d\theta\label{3}
\end{equation}
with $\hat q$ expressed in $p$--representation as:
\begin{eqnarray}
\hat q &= &i(1+p^2) \partial_p= i\partial_\theta,  \nonumber \\
\theta &\equiv& \arctan p \qquad ,  \label{4}
\end{eqnarray}
since $\langle \hat q f \vert g \rangle= \langle  f \vert \hat q g 
\rangle$ by integration by parts,  provided the domain of $\hat q$ is the
set of functions that vanish at the limits of integration $\pm \pi/2$. There are important "self
adjoint  extensions" of $\hat q$. Abbreviating $f(\tan
\theta)=F(\theta)$,  
$\hat q$ becomes an essentially self adjoint operator on the domain \cite{RN}
wherein
\begin{equation}
F(\pi /2)= e^{i \alpha}F(-\pi /2)\label{5}
\end{equation}
It then follows from eqs (\ref{4}) and (\ref{5}) that the position eigenfunctions
 in momentum space are $e^{i q_\alpha \theta}/\sqrt{\pi} $ where $q_\alpha$ are
lattice points
\begin{equation}
q_\alpha= 2\; k +\alpha /\pi \qquad \mbox{\rm{$k$ integer}} \label{6}
\end{equation}
and $0\leq \alpha < 2\;\pi$. The choices of $\alpha$ define physically
indistinguishable bases of the Hilbert space (corresponding to the inability
to localize). Thus any problem governed by KMM dynamics is posessed of
 a U(1) symmetry corresponding to these equivalent choices. \\
  To familiarize the reader with the consequences of this kind of effective
quantum mechanics,  consider the square well problem
\begin{equation}
{\hat p}^2 \Psi = 2\,  m\,  E\,  \Psi \qquad \mbox{\rm with }\Psi(0)=\Psi(L)=0 \label{7} \qquad . 
\end{equation}
One may then work in $q$--representation wherein
\begin{equation}
\hat p=\tan (-i\partial_q) \label{8}
\end{equation}
acting on the subspace of functions of wave lengths greater than 4 (in conformity
with the domain of convergence of the power series of the tangent around
zero \cite{KMM}). It follows that 
$\Psi$ is an eigenfunction of
$\partial_q^2$ given by
$C\,  \sin(n\, \pi\, q/L)$ of eigenvalue $2\, m\, E = \tan^2(n\, \pi/L)$. The spectrum
cuts off at $n=[L/2] $ (i. e. excluding wavelengths shorter than 4). The bracket
$[x]$ symbolizes the integer part of $x$. Note that the problem is rather ill
defined in that $x=L$ is not a physically legitimate concept in that some fuzziness
is always required by eqs (\ref{1}) and (\ref{2}).\\ It is to be expected that
radical effects of the like will emerge in the black hole problem once one tries
to cram a mode too close to the origin. This is our motivation. In this we
recommend \cite{KMM} for a careful discussion of physical states,  maximally
localized states and the critical r\^ole of wave length 4.
\section{Application to the black hole problem} 
For the model of black hole evaporation we take a Schwarzchild black hole,  work
in EF coordinates and neglect the centrifugal barrier that sends  
low frequency($\leq \omega$) outgoing $s$ waves back into the singularity($\omega \leq
(\beta_{H})^{-1}$). These outgoing $s$ waves for $\omega \geq (\beta_{H})^{-1}$
are modes $\psi$ of the form $e^{-i \omega v}  \chi_{\omega}(r)$ where in the
conventional theory one has \cite{BMPS}
\begin{eqnarray}
\label{eq-8}
(1 - 2 M/r) \partial_{r} \psi &=&  -2 \partial_v \psi \qquad ,\label{9a}\\ 
(1 - 2 M/r) \partial_{r} \chi_{\omega} &=&  2 i \omega
\chi_{\omega}\qquad ,\label{9b}
\end{eqnarray}
taken together with the Unruh--Jacobson boundary condition reexpressed by DR in
the form $ p > 0$\footnote{ Recall p is the energy near the horizon , 
$\partial_{x}$ being light like. So $p > 0$ is the restriction to positive
 energy modes near the horizon ($ r- 2 M < 2 M $) and these are
 the modes that give rise to steady state radiation \cite{BMPS}. }. 
 The interesting physics encoded in equation eq. (\ref{9b}) is near the
 horizon where it reduces to
 \begin{equation}
     x \partial_{x} \chi_{\Omega}(x)=  i \; x\;\hat p \chi_{\Omega}(x)=  
4\, M\, i\, \omega \chi_{\Omega}(x)\equiv i\, \Omega \chi_{\Omega}(x)\qquad ,\label{10}
\end{equation}
where $\chi_{\omega}$ is relabeled as $\chi_{\Omega}=\beta_H\omega/(2\;\pi)$ is the dimensionless
frequency.
We have identified $\hat p$ to $-i\;\partial_{x}$ in $x$--representation (where $x= r -2
M)$. 
Thus in eq. (\ref{10}) the units of length drops out of both sides. It will be reinstated
subsquently. In order to
have a complete set of states,  so as
 to fulfill the boundary condition of continuity as the outgoing mode crosses
the star's surface to emerge into the exterior Schwarzchild space,  both signs of
$\omega$ must occur in the linear combination of these positive  energy ($p>0$)
modes. One calls these the ``in" modes,  the basis of second  
quantization in the distant past. The ``out" modes,  those counted by the distant Schwarzschild observer
in the future have a fixed sign of $\omega$,  since there the space is flat. It is
the mixture of positive and negative $\omega$ which characterizes positive
energy modes near the horizon that encodes Hawking evaporation. See ref.
\cite{BMPS} for explanations. 
 
We now propose to adopt KMM for this problem and use eq. (\ref{8}). The
ensuing equation is difficult, to say the least,  so we go over to $p$--representation.
 In this we are (consciously) cavalier in that boundary terms may
get in the way. They do and we shall have more to say on this problem in
 due course. From eq. (\ref{10}),  we then have
\begin{equation}
\label{11}   (1+p^2) \partial_p(p \tilde{\chi}_{\Omega}(p))=  - i 
\Omega \tilde{\chi}_{\Omega}(p)
\end{equation}
or
\begin{equation}
\label{12}
\partial_{\theta}(\tan(\theta) \Phi_{\Omega}(\theta))=  - i \Omega
 \Phi_{\Omega}(\theta)
\end{equation}
where
\begin{eqnarray*}
\Phi_{\Omega}(\theta)=  \tilde{\chi}_{\Omega}(\tan(\theta))
\end{eqnarray*}
The solution is 
\begin{equation}
\label{13}
\Phi_\Omega(\theta)=  A_\Omega [(\sin\theta)^{-i \Omega}\;\cot \theta]\, 
\Theta(\theta)
\end{equation}
where we have applied the vacuum condition $p>0$, hence $\theta>0$. 
Taking the Fourier transform gives  
 \begin{equation}
\label{14}
\chi_{\Omega}(x)=  A_\Omega \int_0^{\pi/2}  e^{i \theta x}\,
\Phi_{\Omega}(\theta)\,d\theta
\end{equation}
The constant $A_\Omega$ will be fixed subsequently. We repeat that $x$ and $p$ are
 non dimensionalised by the unit $\mu$ (
i.e  $x = \mu X $ where all dimensionful quantities are in Planckian units. The
integral (\ref{14}) is feasible and we record the answer in terms 
of the  Beta--function and the hypergeometric
$_2 F_1$:
\begin{eqnarray}
\chi_{\omega}(x) & = &  A_\Omega \,  2^{i \Omega } e^{\pi \Omega/2} \left\{          
B(i\Omega /2+x /2, -i\Omega) \frac{ x}{ x- i \Omega} 
\right.  \nonumber\\
& & \left. -
\frac{2\, e^{i \pi x/2} e^{-\pi \Omega/2}}   {(i\Omega+x+2)(i\Omega+x)}
\, _2F_1(i\Omega/2+x/2, 1+i \Omega;i\Omega/2+x/2+2 ;-1) \right\} \nonumber \\
&&
\label{15}
\end{eqnarray} 
But it is more informative to examine the properties of the integral (\ref{14})
directly. The salient features are:
\begin{enumerate}
\item  The function $\chi_{\Omega}(x)$ defined by eq. (\ref{14}) is not a solution of $ x\;
\hat p \, \chi_{\Omega}= \Omega \, \chi_{\Omega}$ ( with $\hat p =  \tan(-i \partial_{x}$). In
performing    the usual integration by parts one picks up a boundary  
contribution
at $\theta= \pi/2$ ( i.e. $p= \infty$ ),  so that $(x\;
\hat p-\Omega)\chi_{\Omega}(x)=   A_\Omega e^{i \pi x/2}$,  a term which oscillates on the
scale of non--locality. Recall here that $x$ is an affine parameter on the outgoing
geodesic near the horizon. Thus its average effect on this trajectory vanishes. 
We return to further
discussion of this point after the other features are presented. 
\item Asymptotically ($x>>\Omega$) one has $\chi_{\Omega}(x) \rightarrow A_\Omega \,
 x^{i \Omega}$ as is easily seen by carrying out  steepest descents on eq.
(\ref{14}) or using Beta function properties. The width of the saddle at $\theta=
\Omega/x $ is of order $O(\Omega/x^2)$ assuring the validity of the
 estimate. This tallies with the direct analysis of $ \tan(-i\partial_x)\chi = 
(\Omega/x )\chi$ obtained by expanding $ \tan(-i\partial_x) $ in powers  
 of
$\partial_x$. The  constant $A_\Omega$ is then chosen to conform to the conventional
norm of the Klein Gordon current (see ref.\cite{BMPS}). Following the argument of
Jacobson,  Hawking radiation then follows. 
\item  The behaviour of $\chi_{\Omega}(x)$ for $-\Omega< x < \Omega$ confronts
the transplanckian problem neatly. At $x =  \Omega $ the solution changes in
character from the would be rapidly oscillating solution ($x^{i\Omega}$) to a slowly varying 
function. Indeed one calculates directly 
 $\chi_\Omega(0)= ( i /\Omega)A_\Omega$,  and for example $\chi_\Omega(1) - \chi_
\Omega(-1)=  -\left(2/(\Omega+i)\right)A_\Omega$; typically,  $\Omega= O(1)$. \\
  This changeover of behaviour is to be expected,  since along
the characteristic defined by the null outgoing geodesic the point $x= \Omega$
is where $p= 1$ and this is where the important manifestation of nonlocality
sets in,  according to eqs (\ref{1}) and (\ref{10}) . Any tight packet
must then spread over a distance at least comparable  
to the unit of non locality. Hence all sense of hugging the horizon is lost. 
Indeed ,  in a sense we
are really transcending the rules in this small region,  in that equations
\ref{1} and \ref{2} are taken to define an  
 effective theory which allows
one to extrapolate down to the scale of non
 locality,  but not beyond. (In this we are deliberately ambigous because we
 are not sure whether the relevant scale is $x\simeq 1$ or $p\simeq 1$ or something
else. Practically speaking,  since $\Omega= O(1)$,  the question is not of
 much importance,  but conceptually it should be cleared up).
\item   In the  conventional theory Klein--Gordon current conservation is easily
confirmed and is encoded in p representation through the integral
\[  (2 \pi)^{-1} \int_0^{\infty} (dp/p) \, p^{-i(\Omega-\Omega^{\prime})} = 
\delta(\Omega-\Omega^{\prime})  \]
This integral now becomes (see eq. (\ref{21}) below)
 \[   (2 \pi)^{-1} \int_0^{\pi/2} d\theta \cos(\theta)
(\sin(\theta))^{-i(\Omega-\Omega^{\prime})-1} \] which is not a $\delta$
function. 
This non orthogonality has profound repercussions on conservation theorems 
encountered in the evolution of wave packets. We report on this below. 
\end{enumerate}

  Let us now return to what appears to be somewhat of a snag in these
results,  point [1] above. As mentioned the difficulty comes from the upper limit
in eq. (\ref{14}), $p= \infty$ and this is really pushing the mechanism of non--locality
 well beyond what ought to be the range of its validity. Some
regulator is therefore to be called upon. We mention two possibilities,  each of  
which is not unattractive,  but there are surely more. One can expect these will
be revealed upon investigation of the fundamental theory. \\
  The first of these is related to the ambiguity of a choice of origin,  
the equivalent representations labelled by $\alpha$ (see eqs (\ref{5})  
and (\ref{6}). 
We have couched our (cavalier) treatment in terms of the continuum $x$ (which
would appear legitimate since $\chi_\Omega (p) \rightarrow 0$
as $p \rightarrow \infty$ but we have run into trouble because the operator 
$\hat x\, \hat  p$ transforms $\chi_\Omega (p)$ into a function that does not vanish at the
limits. Therefore it would seem mandatory to average $\hat x\;\hat p\;
 \chi_\Omega (p)$ over $\alpha$. Now notice that in point [1],  the term $A_\Omega e^{i
(\pi /2) x}$ acts as a source term for the operator $\hat x\;\hat p - \Omega$. The average of
this source over 4 units of non locality is zero. As we mentioned previously,  
states of wavelength 4 and smaller are to be excluded from physical states. 
Thus the oscillations $e^{i (\pi /2) x}$ of $(\hat x\;\hat p - 
\Omega)\chi_\Omega $ are "unphysical". Only the average over at least 4 length
units is meaningful and the average over 4 does vanish. Alternatively,  
 one may project $(\hat x\;\hat p - \Omega)\chi_\Omega $ on the physical states $ \vert
 \xi >$ of KMM \cite{KMM}. Since these packets have components of wavelength $>$
4,  this projection will vanish. In this sense the averaging process is mandatory,  
and indeed one must anticipate that some averaging of the
 modes $\chi_\Omega(x)$ is necessary to give the theory sense. Once more,  at the
 present stage,  it is difficult to give and exact recipe for averaging. 
Along theses lines,  an alternative procedure is also
possible,  regularizing so as to exclude the component $ p= \infty$ from $(\hat x\;\hat p -
\Omega)\chi$. One may imagine that this will occur from
 an effective action which issues from the same fundamental theory from which the term
$p^2/\mu^2$ arises in the effective commutator and which is a
manifestation of the
 same non locality. A very simple recipe is to add to the equation of motion an
infinitesimal term, $i\,\epsilon\, p^2$,  to give:
\begin{equation}
(\hat x\;\hat p +i\,\epsilon\, \hat p^2)\chi_\Omega= \Omega \chi_\Omega,  \qquad . \label{15}
\end{equation}
At small $p$ (large $x$ in wave packets)
the correction is negligible and at large $p$ ( small $x$) it
 takes care of the problem of the upper limit,  so as to rid one of the $e^{i (\pi
/2) x}$ oscillating source. One easily checks that $\chi_\Omega$ now becomes $A_\Omega
[(\sin \theta)^{i \Omega}/ \tan \theta ](\cos  \theta)^{i\,\epsilon}
$. The term $(\cos\theta)^{i\,\epsilon} $ is sufficient to
eliminate the spurious source term (in the sense of distribution theory). Clearly the two methods are carrying similar
messages; an average over a few units of non locality is necessary. 

We now turn to our analysis of the evolution of wave packets. 
As a preliminary we first consider conservation of current and charge. 
This is carried out, following Noether, in conventional fashion in momentum 
representation. Using eq. (\ref{12}) one constructs the current $(j^v,j^\theta)$ where
\begin{eqnarray}
j^v&=&\frac 12 \left[\Psi^*(\theta)\,(\tan \theta \, \Xi(\theta))+(\tan \theta \, \Psi(\theta))^* \Xi(\theta)\right]\label{18a}
\\ j^\theta&=&(\tan \theta \, \Psi(\theta))^* (\tan \theta \, \Xi(\theta)) \label{18b}
\end{eqnarray}
where $\Psi$ and $\Xi$ are solutions, whence $\partial_v j^v +\partial_\theta j^\theta=0$.\\
  This conserved current, nevertheless, does not lead to a conserved charge
 owing to boundary terms in $\theta$.
The ``would be'' conserved charge is the diagonal element of the general form 
$\langle {\bf \Psi},{\bf \Xi} \rangle$ where
\begin{equation}
\langle {\bf \Psi},{\bf \Xi} \rangle=\int_0 ^{\pi/2} \Psi^*(\theta)\,(\tan \theta) \Xi(\theta) \; d\theta \label{19}.
\end{equation}
From eq.(\ref{12}) one then has
\begin{equation}
\frac d{dv }\langle {\bf \Psi},{\bf \Xi} \rangle=\Psi^*(\theta)\; \Xi(\theta)\; \tan^2 \theta \vert\, _ 0 ^{\pi /2}\label{20}
\end{equation}
As an example previously cited, the modes are not orthogonal
\begin{equation}
\langle \phi_\Omega, \phi_{\Omega ^\prime} \rangle=A^*_{\Omega^\prime} 
A_\Omega [\pi \delta(\Omega-\Omega^\prime)-i\,\mbox{\rm P}\; \frac
{e^{i(\Omega^\prime-\Omega)}}{\Omega^\prime-\Omega}]\label{21}
\end{equation}
where we have written $\phi_\Omega=e^{-i \Omega V}\Phi_\Omega(\theta)$ 
(and $V\equiv v/4M)$.\\
  The principal value term in (\ref{21}) comes from the upper limit of 
the $\theta$ integral in (\ref{19}), i.e. it is a high momentum effect. Unlike point 
(1) above, there is no argument on hand to eliminate this effect since there is no
spatial averaging procedure available. In fact, any cut--off of the wavelength 
will yield such effects.\\
  In $x$--representation the corresponding charge is:
\begin{equation}
\langle {\bf \Psi},{\bf \Xi} \rangle=\frac 12 \int_{-\infty}^\infty 
 [\psi^*(x) \tan(- i \partial_x)  \xi(x) - \xi (x)
\tan(-i \partial_x) \psi^*(x)]\; dx\label{22}
\end{equation}
where $\psi(x)=\int_0^{\pi /2}e^{i \theta x}\Psi(\theta)\, d\theta$ and similarly for $\xi(x)$ and $\Xi(\theta)$.
 The limits on $x$ are chosen to be $\pm \infty$ since we are interested in wave packets
 of finite extent. Such packets are centered asymptotically on $u=v-4M \ln \vert
 x\vert=const$ i.e. in the region where the dominant components of $p$ are $p<1$ 
(thus $\theta<<\pi/2$). In this region, $\tan(\frac 1 i \partial_x)$ acts like 
$(\frac 1 i \partial_x)$ and conventional charge conservation obtains. When $v$ 
is sufficently early the center of the packet enters into the region of non--locality 
($\vert x\vert < \Omega $ or $p>1$) and the above effects of charge non conservation set in.\\
  We have performed numerical computations on a packet of the form
\begin{eqnarray}
{\cal F}(V,\theta)&=&\int e^{-(\Omega-\Omega_H)^2/\sigma^2}\frac {A(\Omega_H)}{A(\Omega)}\phi_
\Omega(V, \theta) \;d\Omega \nonumber \\&=& {A(\Omega_H)}e^{-i \Omega_H (V+\ln\sin\theta)}
\cot\theta \, e^{-\frac{\sigma^2} 2(V+\ln \sin \theta)^2}\label{23}
\end{eqnarray}
centered on frequency $\omega_H (=\Omega_H /4M )$. In configuration space it is asymptotically centered on the 
trajectory $ u=const$ (the value of the latter is irrelevant) where $u=v-4M\ln |x|$. We
 present in the accompanying figure the evolution of the packet in configuration space,
 given by $f(V,x)=\int_0 ^{\pi /2}e^{i\theta x}{\cal F}(V, \theta) d\theta$. We now discuss
 its qualitative features.\\
  In the asymptotic region, the conventional non spreading packet, centered on 
$u=const$ obtains.For $V<V_H (\equiv
e^{\Omega_H})$ (i.e. for values of $V$ which are earlier than that which corresponds 
to $x=\Omega_H$ along the asymptotic trajectory), the center of the trajectory deviates 
from $u=const$ and bends in towards the horizon $(x=0)$. It crosses the horizon and in the 
region $x<-\Omega_H$ becomes the classical trajectory of the Hawking "partner" which falls 
into the singularity. In the non local region $\vert x\vert < \Omega_H$, the classical trajectory 
is not of quantitative significance since the packet spreads in $x$ over the region of non locality.\\
  These features are qualitatively analyzed as in ref \cite{BMPS} by a saddle point 
calculation of $ f(V,x)$ or alternatively by the method of characteristics for the trajectory 
of the center of the packet. One finds for $p(V)$ the equation $p^2=e^{-2(V-V_H)}/(1-e^{-2(V-V_H)})$.
 Thus the conventional behavior obtains for $V>V_H$ whilst for $V<V_H$, $p$ behaves like $\pm[V-V_H]^{1/2}$.
 It is to be noted that there is no extrapolation to the past for $V<V_H$ (i.e. the notion of a 
usual causally behaved trajectory stops). Correspondingly $x\simeq \Omega /p$ for $(V-V_H)>>1$
 and $x \simeq \Omega/p^3$ for $V-V_H\simeq 0$. We again emphasize that this classical movement 
in the region $\vert x\vert <\Omega_H$ is without quantitative significance since the width in 
$x$ in this region is $O(\Omega_H)$.\\
  Most interesting is the evolution of the non conserved charge.
 One can calculate explicitly
 from eq. (\ref{22}) with $\tilde \psi$ and $\tilde \Xi$ given by $\tilde f(V,x)$ 
(the Fourier transform of (\ref{23}))
\begin{eqnarray}
Q(V)&=&{2 \pi }\vert A(\Omega_H)\vert^2 \int_0 ^{\pi/2}e^{-\sigma^2(V+\ln \sin 
\theta)^2}\cot \theta \, d\theta \label{24}\\ 
&=&{2 \pi }\vert A(\Omega_H)\vert^2\frac{\sqrt \pi}{2\sigma}[1+\mbox{\rm Erf} (\sigma
V)]\label{25}
\end{eqnarray} 
where $\mbox{\rm Erf}$ is the error function and we have chosen $V_H=0$. Thus, for large $V$, 
$Q(V)$ is $[const + O(e^{-\sigma^2 V^2})]$ whereas for $V<0$, $Q$ tends to zero like
 $e^{-\sigma^2 V^2}$. The behavior changes around $V=0$ corresponding to a saddle in 
(\ref{24}) at $\ln \sin \theta^*=-V$) with $\theta^*$ near to $\pi/2$ i.e. $p\geq 1$. 
Thus the charge diminishes rapidly within the non local region. Read forward in time, 
this means that the "stuff" in the packet is produced in this region. 
\section{Concluding remarks}
  From these various considerations one obtains an interesting interpretation
wherein the region where non locality plays an active role : ($-
\Omega_H < x < \Omega_H$) is the zone of interaction of the Hawking photon mode with
reservoir modes. Going forward in time ,  this region which straddles  
the
horizon over a scale $ \Omega_H$ in units of non local length boils off
 a pair which for $|x| > \Omega_H$ looks like that of the conventional theory. It
is  "solicited" by the collapse,  this latter being encoded in the boundary
conditions characterizing the Unruh vacuum. Clearly,  much more work  
is required
in order to see just how this encoding is dynamically realized (perhaps
combining Unruh's analysis of the collapsing shell with the present
considerations or perhaps calling upon 't Hooft's scattering mechanism between 
incoming and outgoing degrees of freedom \cite{tHooft}). 

It is highly significant that the theory based on the effective commutator (\ref{1}) 
is non unitary as well as non
causal  in the usual sense of the words. Whereas the latter was to be anticipated at
 the outset, the former has arisen as a consequence. In the approach of this paper, curing the
transplanckian problem is inevitably asssociated with non unitarity in that the sector of
Hilbert space describing the states of the light modes giving rise to Hawking radiation
is not complete.
This should come as no surprise in string theory since then the scattering matrix defined by
the zero--mass sector is not unitary either. One will produce quanta of massy modes once
energies and momenta are high enough. Apparently this effect is encoded in eq. (\ref{1}).

  One may conjecture that the loss of Bekenstein--Hawking radiation
entropy is due to this boiling off  from the reservoir. It occurs in a "zone of
ignorance" composed of $\Omega$ units of length where $\Omega= 
\beta_{H} \omega /(2 \pi)$. Could it be that each unit corresponds to a volume $
2 \pi$ of "ignorance" due to the inability to localize in that region?The
volume $ 2 \pi$ is of course the group volume of the $ U(1)$ of Kempf. Much work
is required to substantiate this conjecture,  but it is remarkable that the
length scale over which a given Hawking mode "gets lost"
 due to its interaction with the reservoir is $\Omega$ and not 1. And of course
this is what incites one to make the conjecture.

  Our approach then invites a series of problems and still further conjectures.
First of all what is the physical interpretation of eqs (\ref{1}) and (\ref{2}). 
The problem comes in two parts, general and particular to the situation at hand.\\
  As to the former, it it clear from the string theoretic model that the 
modification that sets in for the quantum description of the zero mass states of 
momentum $p>\mu$ (where $\mu\simeq \sqrt {tension}$) is related to the fact that 
such modes can dissipate into the reservoir of the higher mass modes of the string. 
The density of states of these latter is sufficient to create this dissipation. 
As Renaud Parentani has suggested to us, the situation is reminiscent of the Hagedorn 
temperature wherein energy that is poured into the zero mass modes gets redistributed 
into the higher mass modes. Thus increasing the energy of the former does not result 
in an increase of their temperature (their mean energy) once the energy becomes 
comparable with $\mu$. Similarly were one to measure $Q$ with zero mass modes 
with increasing precision one would be frustrated since increasing the energy of 
these modes would be of no avail. Hence a minimum of precision is reached which 
in fact is of the order of the inverse Hagedorn temperature.

  Another interesting point in this same vein is the conjecture that there 
is a sort of quantum Nyquist theorem; see for example \cite{stat}. Dissipation of a given set of modes 
into a reservoir implies that such modes " jiggle" due to their recoil induced by the 
dissipating interactions. Then the thermal noise (the jiggle) is due to the vacuum 
fluctuations of the reservoir and the drag term will be the rate of dissipation. 
Equation (\ref{1}) would then be viewed as a time dependent commutator averaged over the dissipation time
and the reservoir states.\\
  Accepting the reservoir interpretation, which seems inevitable, what 
then is the nature of the reservoir in our black hole problem. Does it concern 
vacuum fluctuations in general (usually swept under the rug due to the 
renormalizability of field theory relevant to particle physics, but highly relevant 
in production from horizons)? If so, then the problem would be related to current
 efforts to confront quantum gravity such as string or M theory. Or is the reservoir 
specific to the black hole (or horizons in general)? Serge Massar has raised the 
possibility that the reservoir modes are vibrations of the membrane of the so--called
 membrane paradigm.

It is far from clear that one can develop a satisfactory phenomenology
 without a complete understanding of the reservoir. If it is possible, the 
theory would occupy some middle ground between dynamics and thermodynamics 
prior to a full understanding which we all believe will require the correct theory of quantum gravity.

\noindent{\bf Acknowledgements}

  RB wishes to express his gratitude to Gianpiero Mangano for introducing 
him to the KMM theory. In addition, very special thanks are due to Achim Kempf who 
read the initial manuscript with great care and then suggested several improvements 
which have made our presentation much clearer. Discussions with Serge Massar 
and Renaud Parentani have been of invaluable help in the elaboration of physical
 interpretation and in our understanding of the motion of packets. We thank them warmly.\\

\vfill
\newpage
\noindent {\bf Figure caption: }\\
  The evolution of the wave packet of eq.(\ref{23}) is depicted. 
The parameters are $\Omega_H=1,\, \sigma=1, \, V_H=0$.
The coordinates are advanced Eddington--Finkelstein ($V$ and $x$ where $V=v/4M$ and $x=(r-2M)/\mu$).
Ingoing null geodesics are straight lines plotted on a 45--degree slant. 
 The center of the packet is indicated by the dashed line. It is 
classical trajectory for $\vert x\vert >1$ (i.e. $u=constant$; such an outgoing null geodesic is plotted on the bottom of the
figure). The shaded region  indicates the spread. For each $x$ the spread in $V$ is constant $(\simeq 4)$ 
and this gives the large spread in $x$ at $V=V_H (=0)$, the time about 
which the packet enters into the region of non locality.\\

\begin{thebibliography}{999}

\bibitem{Hawk} S.W. Hawking, Nature {\bf 248} (1974) 30; Commun. Math. Phys.,{\bf
43} (1975) 199.
\bibitem{BMPS} R. Brout, S. Massar, R. Parentani, Ph. Spindel,
Phys. Rev. {\bf D 52} (1995) 4559.
\bibitem{Unruh}W. Unruh, Phys. Rev. Lett. {\bf 21} (1981) 1351; 
Phys. Rev. {\bf D 51} (1995) 2827.
\bibitem{DR} T. Damour, R. Ruffini; Phys. Rev. {\bf D 14} (1976) 332.
\bibitem{KMM} A. Kempf, G. Mangano, R.B. Mann, Phys. Rev. {\bf D 52} (1995) 1108.
\bibitem{Kempf} A. Kempf, G. Mangano, Phys. Rev. {\bf D 55} (1997) 7909.
\bibitem{Bek} J. Bekenstein, Phys. Rev. {\bf D 7} (1973) 2333.
\bibitem{RN} F. Riez, B. Sz.--Nagy, {\sl Le\c cons d'Analyse Fonctionnelle},
Gauthier--Villars (1975), Chap. {\bf VIII}, $\mbox{\rm n}^{\mbox{\rm o}}\ 119$.  
\bibitem{} A. Kempf, Europhysics Lett. {\bf 40} (1997) 257.
\bibitem{Jacobson} T. Jacobson, Phys. Rev. {\bf D 48} (1993) 728.
\bibitem{tHooft} G. 't Hooft, Int. J. Mod. Phys. {\bf A 11} (1996) 4623.
\bibitem{stat} C. Kittel, {\sl Elementary Statistical Physics}, New York . London . Sydney,
John Wiley \& Sons, Inc. (1958), section {\bf 29}.
\end{thebibliography}
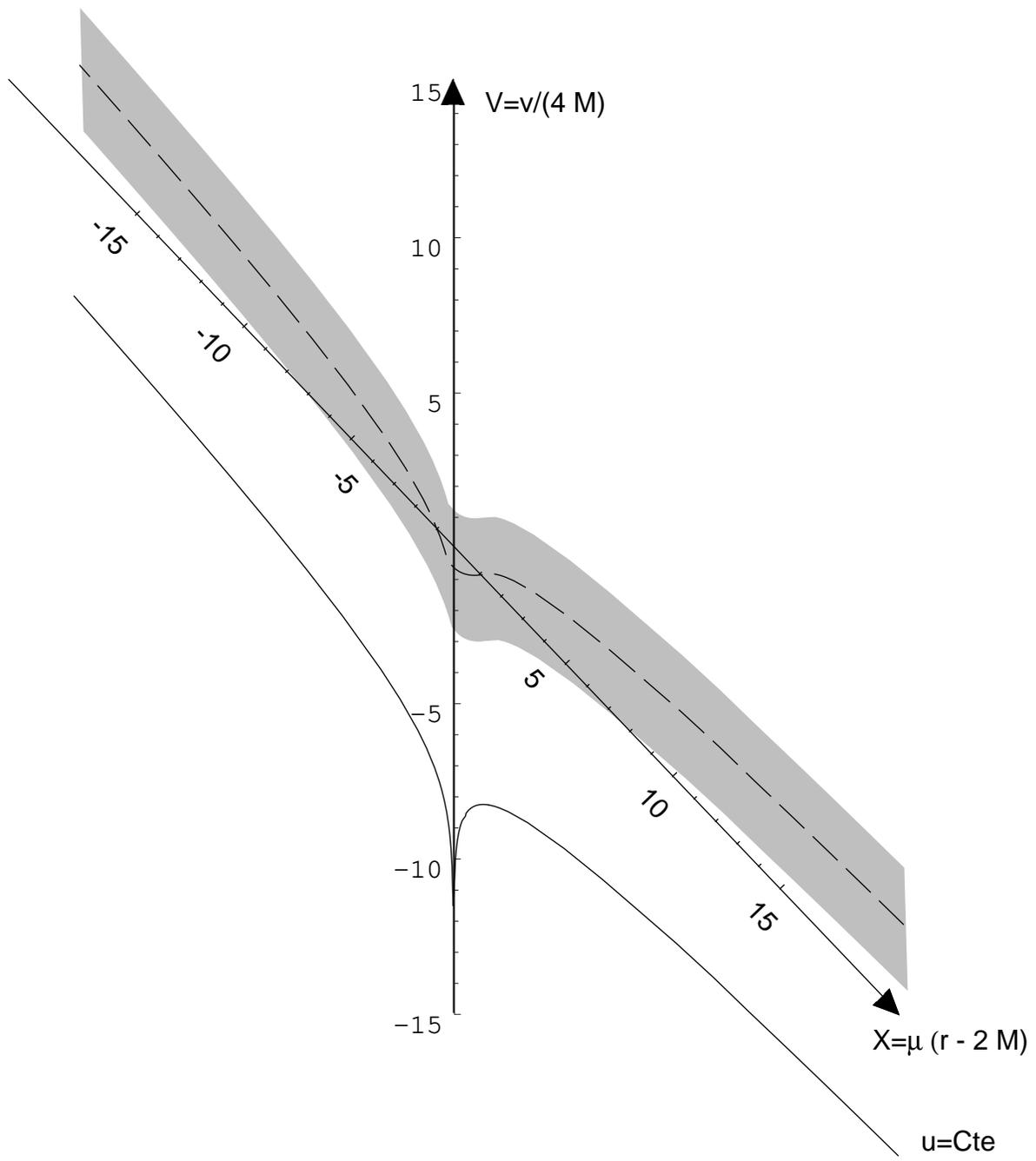
\end{document}